\documentclass[%
 notitlepage,
 reprint,
 superscriptaddress,
 amsmath,amssymb,
 aps,
]{revtex4-1}

\usepackage{lipsum}
\usepackage{graphicx}
\usepackage{dcolumn}
\usepackage{bm}

\usepackage{epstopdf}

\newcommand{\bra}[1]{\langle #1 \vert}
\newcommand{\ket}[1]{\vert #1 \rangle}

\begin{document}

\preprint{APS/123-QED}

\title{Interplay between speed and fidelity in off-resonant quantum-state transfer protocols}

\author{Guilherme M. A. Almeida}
\email{gmaalmeida.phys@gmail.com}
\affiliation{%
 Instituto de F\'{i}sica, Universidade Federal de Alagoas, 57072-900 Macei\'{o}, AL, Brazil
}%

\date{\today}

\begin{abstract}
An arbitrary qubit can be transmitted through a spin chain
by perturbatively coupling both communicating parties to it.
These so-called weak-coupling models rely on
effective Rabi oscillations between the outer spins,
yielding nearly maximum fidelity while offering great 
resilience against disorder with the cost of having long transfer times. 
Considering that framework, here we address a 1D non-symmetric channel connecting two spins, one placed at each end of it.
Given any pattern of nearest-neighbor coupling strengths, 
we obtain an analytical expression that accounts for 
the effective long-range interaction between them and study
the interplay between transfer time and fidelity.
Furthermore, we show that homogeneous channels
provide the best speed-fidelity tradeoff.

%



\end{abstract}

\maketitle


\section{\label{sec1}Introduction}

Spin chains have been extensively exploited for many quantum information processing tasks such as the 
transmission of quantum states \cite{bose03,christandl04,albanese04,plenio04,niko04,*niko04epl,osborne04,wojcik05,wojcik07,li05,kuznetsova08,huo08,liu08, gualdi08,wang09, banchi10, *banchi11,*apollaro12,lorenzo13,lorenzo15, almeida16, almeida17-2} 
and the creation/distribution of entanglement between distant sites
\cite{amico04, apollaro06,*plastina07,*apollaro08,venuti07, giampaolo09, *giampaolo10, gualdi11, estarellas17, *estarellas17scirep, almeida17-1}
(cf. Refs. \cite{apollarorev, kayrev,nikobook} for reviews on the subject). 
As it was put foward in \cite{bose03}, the scheme
is based on wiring up different quantum processing units via spin-spin exchange interactions and let them evolve following
the natural Hamiltonian dynamics. 
This means that 
the Hamiltonian can be engineered
in order to avoid dynamical manipulation during 
the quantum communication task. 


In standard single-qubit quantum-state transfer (QST) protocols \cite{bose03}, 
the sender, say Alice, sends an arbitrary state through the channel
and Bob's only role is to retrieve it at some prescribed time.
The interplay between transfer fidelity and speed 
will be dictated by way the chain is manufactured.
Perfect QST can be achieved in fully-engineered chains \cite{christandl04, plenio04,niko04,feder06}. 
In order to bypass some possible practical issues \cite{dechiara05, zwick11, *zwick12} concerning 
tuning the entire set of couplings within the chain,
several other schemes were put forward.
For instance, by locally adjusting the end bonds of the chain, it is possible
to perform ballistic transfer through an arbitrarily long chain  \cite{banchi10}.
One may also apply 
strong magnetic fields
near the communicating parties in order to energetically detach them from the channel \cite{plastina07, lorenzo13, paganelli13}.    
%

Here, in particular, we deal with a similar class of protocols based on setting \textit{weak} couplings 
between the end spins 
and the rest of the chain (bulk) \cite{wojcik05,wojcik07, li05,kuznetsova08,huo08, almeida16} 
in order to effectively 
span a reduced subspace involving the sender and receiver only, up to leading order. 
That kind of configuration has also been addressed for the sake of 
generating long-distance entanglement \cite{venuti07,giampaolo09,oh10}.
In the QST context,
it entails 
nearly perfect transfer, though requiring very long times. Taking
$a$ as the (coupling) perturbation parameter, the transfer time scales as $O(a^{-2})$ for a channel featuring an even number of sites \cite{wojcik07}
operating in the Rabi-like (that is, two-level) off-resonant regime,
where the frequency of both outer spins does not match any of the natural frequencies of the channel.
In this work, 
we explore in detail the inner workings of that class of QST protocols
and discuss the speed-fidelity balance
for channels with \textit{arbitrary} (non-symmetric) couplings. 
We derive a simple, exact formula that
accounts for the end-to-end effective coupling strength
as a function of the coupling sequence of the channel,
up to second order perturbation theory. By using it, we investigate
the speed-fidelity cost
for a variety of configurations and
show that uniform channels \cite{wojcik05} are optimal in respect to time
in the weak-coupling regime.


Next, in Sec. \ref{sec2}, we 
introduce the $XX$ spin Hamiltonian with 
arbitrary coupling strengths and work out the perturbation approach in
Sec. \ref{sec3}. Then, in Sec. \ref{sec4},
we derive 
a relationship between the coupling pattern of the channel
and its related QST time in the Rabi-like regime.
In Sec. \ref{sec5}, we 
discuss the response of the end-to-end correlation amplitude 
to the channel coupling scheme.
Our conclusions are drawn in Sec. \ref{sec6}.



\section{\label{sec2}Spin Hamiltonian}

We consider $XX$ spin-$1/2$ chains with open boundaries featuring $N+2$ sites, 
with $N$ (even) being the length of the channel (that is, the bulk of the chain) and the remaining spins acting as the sender/receiver parties which we denote by $S$ and $R$.
They are connected, respectively, to each end of the channel, sites $1$ and $N$. 
The Hamiltonian of the system is expressed as $\hat{H} =\hat{H}_{\mathrm{ch}}+\hat{H}_{I}$,
with ($\hbar = 1$)
\begin{align} \label{Hchannel}
\hat{H}_{\mathrm{ch}} &=\sum_{i=1}^{N-1}\dfrac{J_{i}}{2}(\hat{\sigma}_{i}^{x}\hat{\sigma}_{i+1}^{x}
+\hat{\sigma}_{i}^{y}\hat{\sigma}_{i+1}^{y}), \\
\hat{H}_{I} & = 
\dfrac{a_{S}}{2}(\hat{\sigma}_{S}^{x}\hat{\sigma}_{1}^{x}
+\hat{\sigma}_{S}^{y}\hat{\sigma}_{1}^{y}) +
\dfrac{a_{R}}{2}( \hat{\sigma}_{N}^{x}\hat{\sigma}_{R}^{x}
+\hat{\sigma}_{N}^{y}\hat{\sigma}_{R}^{y}), \label{Hint}
\end{align}
where $\hat{\sigma}_{i}^{x,y}$ are the Pauli operators for the $i$-th spin, 
and $J_{i}$ and $a_{S,R}$ are nearest-neighbor spin coupling strengths.
Herein we define $J = J_{\mathrm{max}} \equiv 1$ as the energy unit. 
The local magnetic fields are set uniformly across the chain ($\omega \equiv 0$). 
Since $\hat{H}$ commutes with the total magnetization operator, 
$\left[ \hat{H}, \sum_{i} \hat{\sigma}_{i}^{z}\right] =0$, the Hamiltonian is block diagonal with respect to the number of excitations. 
In this work we only consider single-excitation states which can be expressed by
$\ket{j} \equiv \ket{1}_{j}\ket{\textbf{0}}$, 
namely a spin up located at site $j$  ($j = S,1,2,\ldots,N,R$) with the remaining ones in their ground state $\ket{\textbf{0}} \equiv \ket{00\ldots 0}$.

%
Basically, here we deal with
a generalized version of the weak coupling-based QST, 
with arbitrary channel couplings $\lbrace J_{i} \rbrace$ and $a_{S,R}\ll J$ in general.
Particularly, our discussion 
is focused on the off-resonant two-level dynamical regime, whereby 
spins $S$ and $R$ span their own subspace, up to a second-order approximation,
so that QST occurs via Rabi-like oscillations between them \cite{wojcik07}.  
%

%
\begin{figure}[t!] 
\includegraphics[width=0.45\textwidth]{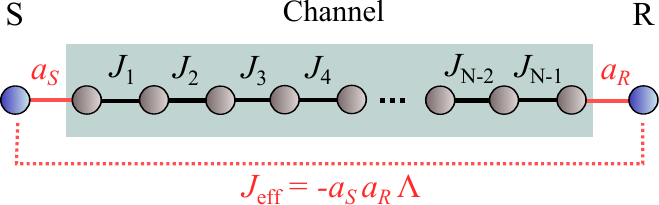}
\caption{\label{fig1} (Color online) Coupling scheme for the $XX$ spin chain described by Hamiltonian (\ref{Hchannel}).
Both end spins denoting the communicating parties $S$ and $R$ are
perturbatively connected, at rates $a_{S}$ and $a_{R}$, respectively, to a channel 
featuring an even number of sites $N$ with
\textit{arbitrary} couplings. Up to second-order perturbation theory, both outer spins eventually develop 
an effective interaction $J_{\mathrm{eff}}$ mediated by the channel through $\Lambda = \Lambda(J_{1},J_{2},\ldots,J_{N-1})$ [cf. Eq. (\ref{omega})], 
which characterizes its strength.
}
\end{figure}

\section{\label{sec3}Effective two-level Hamiltonian}

In general -- without specifying any particular coupling scheme for now -- the basic QST protocol \cite{bose03} consists in the transmission of an arbitrary qubit state $\ket{\phi} = \alpha \ket{0}+\beta \ket{1}$  from one point to another. This can be done by 
initializing the whole chain in $\ket{\psi(0)} = \ket{\phi}_{S}\ket{0}_{1}\ket{0}_{2}\ldots \ket{0}_{R}$ and letting it evolve naturally
through $\ket{\psi(t)} = e^{-i\hat{H}t}\ket{\psi(0)}$. In the ideal scenario,
the final output should be as close as possible to
$\ket{\psi(\tau)} = \ket{0}_{S}\ket{0}_{1}\ket{0}_{2}\ldots \ket{\phi}_{R}$
at some prescribed time $\tau$, so that the receiver can properly retrieve the information.
The corresponding transfer fidelity can thus be 
evaluated by $F_{\phi}(\tau) = \bra{\phi} \rho_{S,R}(\tau) \ket{\phi}$, 
where $\rho_{S,R}$ is obtained by tracing out 
the channel spins ($1$ through $N$),
$\rho_{S,R}(\tau) = \mathrm{Tr}_{\mathrm{channel}}\ket{\psi(\tau)}\bra{\psi(\tau)}$.
Now, the figure of merit of the channel may be evaluated by averaging
the transmission fidelity $F_{\phi}$ over all input states $ \ket{\phi}$ (i.e., over the Bloch sphere such that $|\alpha|^{2}+|\beta|^{2}=1$) which -- given 
the dynamics
is restricted to occur 
in the zero- and one-excitation subspaces -- results in \cite{bose03} 
\begin{equation} \label{eq2}
F(t) =\frac{1}{2} + \frac{|f(t)|}{3} + \frac{|f(t)|^2}{6},
\end{equation}
where $f(t)\equiv \bra{R} e^{-i\hat{H}t}\ket{S}$ is the transition amplitude between spins $S$ and $R$.
This quantity ultimately embodies the quality of the transfer.  

In the case of Rabi-type QST \cite{wojcik05, wojcik07, almeida16}, the goal is to design the chain 
such that a couple of eigenstates having the form $\ket{\lambda^{\pm}} \simeq (\ket{S} \pm \ket{R})/ \sqrt{2}$,
with corresponding energy gap $\delta \lambda$,
is obtained. In this scenario, the transition amplitude takes
\begin{align} \label{eq3}
\vert f(t) \vert & \simeq \frac{C}{2}  \vert e^{-i\delta \lambda t / 2} - e^{i\delta \lambda t / 2} \vert \\ \nonumber
& = C \left| \sin{ \left( \frac{\delta \lambda}{2}t \right)} \right| ,
\end{align}
where $C \equiv 2 |\bra{R}\lambda^{\pm} \rangle \bra{\lambda^{\pm}} S\rangle| \simeq 1$ measures the 
end-to-end correlation amplitude.
The transfer time is thus $\tau  = \pi/\delta \lambda $. The energy gap between both
dominating eigenstates sets the typical energy scale of the transfer. 

While there are many schemes to achieve that kind of behavior \cite{wojcik05, wojcik07, li05, gualdi08, kuznetsova08, almeida16}, 
our goal here is to find out the optimal combination $\lbrace J_{i} \rbrace$
which provides the highest possible $|f(t)|$ with maximum $\delta \lambda$ allowed.
In order to do so, we use a second-order perturbation approach (cf. Refs. \cite{wojcik07, li05}) in $a_{S,R}$  so that we 
can derive $\delta \lambda$ in terms of the spectral resolution of the channel.
The effective interaction between states $\ket{S}$ and $\ket{R}$ is accounted by
\begin{equation} \label{Heff}
H_{\mathrm{eff}} = \begin{pmatrix}
h_{S} & J_{\mathrm{eff}} \\ 
J_{\mathrm{eff}} & h_{R}
\end{pmatrix}
\end{equation}
where
$h_{S(R)} =  \sum_{k}(|v_{k,1(N)}|^{2} / \epsilon_{k})$,
$J_{\mathrm{eff}} = -a_{S}a_{R}\Lambda$, $\Lambda = \sum_{k}(v_{k,1}v_{k,N}^{*}/\epsilon_{k})$, 
$v_{k,1} = \langle 1 \vert \epsilon_{k} \rangle$, and $v_{k,N} = \langle N \vert \epsilon_{k} \rangle$, 
with $\ket{\epsilon_{k}}$ and $\epsilon_{k}$ being, respectively, the 
eigenstates and eigenvalues of the channel Hamiltonian [Eq. (\ref{Hchannel})]. 
Note that the spectrum features particle-hole symmetry, that is $\epsilon_{k} = - \epsilon_{-k}$ and $|v_{k,j}|=|v_{-k,j}|$ for every $k$,
and thus $h_{S} = h_{R}=0$.

The resulting two-level gap then reads $\delta \lambda = 2J_{\mathrm{eff}}$.
According to the above effective description, the time it takes to transfer a quantum state from $S$ to $R$ is
$\tau = \pi/(2a^{2}|\Lambda|)$, with $a = a_{S} = a_{R}$ from now on. Since $a$ must very small in order to
assure the validity of 
Hamiltonian (\ref{Heff}), 
we shall expect long transfer times. 
Given $\lbrace J_{i} \rbrace$, the appropriate value for 
$a$ will depend
upon the spectral properties of the channel, its size, and the required fidelity outcome.
In homogeneous channels, for example, it is enough to set $a\ll J/\sqrt{N}$ \cite{wojcik05}.
There is also the possibility of searching for 
coupling patterns that maximize $\Lambda$ in order to counterbalance 
$a$.
In the following section, we derive an exact 
analytical expression for  $\Lambda$ as a function of  $\lbrace J_{i} \rbrace$. 

\section{\label{sec4}End-to-end interaction as a function of the coupling scheme of the channel}

First, it is convenient to introduce the shorthand notation
($J_{1}, J_{2}, \ldots, J_{N-1}$) to represent the coupling sequence
for a $N$-site channel (see Fig. \ref{fig1}). 
Recall that we consider $N$ to be even throughout the paper.
For instance, $(J_{1})$ and $(J_{1},J_{2}, J_{3})$ 
denote a 2- and a 4-site channel, respectively.
Their corresponding values of $\Lambda$ are easily obtained and read
$\Lambda_{2} \equiv \Lambda(J_{1}) = 1/J_{1}$ and $\Lambda_{4} \equiv \Lambda(J_{1},J_{2},J_{3}) =- J_{2}/(J_{1}J_{3})$ (cf. Appendix \ref{appendix}).
The latter outcome is equivalent to that of a dimer with coupling
$\Lambda_{4}^{-1}$.
Now, consider a $6$-site chain with
weak end couplings, $J_{1}, J_{5} \ll J$, for instance. From the perturbation theory framework
discussed previously, the outer sites should develop an effective coupling
of the form $-J_{1}J_{5}\Lambda'$, 
with $\Lambda' = \Lambda(J_{2},J_{3},J_{4}) = -J_{3}/(J_{2}J_{4})$.
Note we can treat the above configuration as a virtual ($4$-site) chain
with coupling sequence $(J_{1}, \Lambda'^{-1},J_{5})$. This yields
$\Lambda_{6} \equiv \Lambda (J_{1},J_{2},J_{3},J_{4},J_{5}) = J_{2}J_{4}/(J_{1}J_{3}J_{5}) $.
Curiously, this very same result should be obtained regardless of the strength of $J_{1}$ and $J_{5}$ (see Appendix \ref{appendix} for details).
Henceforth, using the same reasoning, $\Lambda$ can be built
for arbitrary (even) $N$ and uniform on-site potential distribution 
following the recursive relation:
$\Lambda_{M ,j}= -(\Lambda_{M,j-1}J_{M-j}J_{M+j})^{-1}$ for $j = 1,2,\ldots, M-1$, with $M  = N/2$,
$\Lambda_{M ,j} \equiv \Lambda(J_{M-j},J_{M-j+1},\ldots, J_{M-1},J_{M},J_{M+1},\ldots, J_{M+j})$, and
$\Lambda_{M,0} \equiv \Lambda(J_{M}) = J_{M}^{-1}$.

By iterating the above rule over and over until $j = M-1$ we get
\begin{equation}\label{omega}
\Lambda_{N }\equiv \Lambda_{M,M-1} = 
(-1)^{M+1}\frac{J_{2}J_{4}J_{6}\cdots J_{N-2}}{J_{1}J_{3}J_{5}J_{7}\cdots J_{N-1}}.
\end{equation} 
%
The above equation comprises the key result of this work (a formal derivation of it is provided in Appendix \ref{appendix}).
By weakly switching on the interaction between spins $S,R$ and the channel,
the effective coupling between the outer ends of the chain becomes
$J_{\mathrm{eff}} = -a^{2}\Lambda_{N}$ [cf. Eq. (\ref{Heff})] up to second order in $a$. 
It is worth stressing that the term $\Lambda_{N}=\sum_{k}(v_{k,1}v_{k,N}^{*})/\epsilon_{k}$ 
as shown above in Eq. (\ref{omega}) is exact by \textit{itself}, although it emerges from
the effective Hamiltonian (\ref{Heff}) which, on its side, 
has been derived using a perturbation approach.
As we are about to see, our result displayed in Eq. (\ref{omega}) is
a handy resource for studying
the trade-off between speed and fidelity 
given any pattern of couplings along the channel.

\section{\label{sec5}Speed versus fidelity}

In the following discussion, 
we will often
treat the whole channel mediating spins $S$ and $R$
as a single dimer coupled by $\Lambda_{N}^{-1}$ for convenience. 
In the light of second-order perturbation theory, the full chain
is equivalent to a $4$-site chain with coupling sequence ($a$,$\Lambda_{N}^{-1}$,$a$).
This simple picture shows that
the outer spins 
no longer have access to precise information over the number of spins within the channel 
as well as its exact coupling sequence.
All they ``see'' is $\Lambda_{N}$ which, 
in turn, can be set in infinitely many ways.
The fidelity thus scales as $F \sim 1-O(a^{2}\Lambda_{N}^{2})$
entailing an almost perfect QST 
at times $\tau \sim (a^{2}|\Lambda_{N}|)^{-1}$.

 
It also deserves notice the fact that $\Lambda_{N}$ contains 
useful information about the channel itself. That gives us insight 
over how large is the gap in the center of the energy band,
based on the picture established above.
First, note in Eq. (\ref{omega}) that 
if a given coupling $J_{i}$ promotes
an increase (or decrease) in $\Lambda_{N}$, the next one, $J_{i+1}$, will do the opposite, and so forth. 
This curious fact explain many of the properties featured in staggered chains
with alternating weak and strong bonds (see e.g., Refs.  \cite{venuti07, huo08, kuznetsova08, almeida16}).
This kind of configuration is known for providing very high fidelities on the one hand, even for moderately
distorted couplings, and very long QST times on the other hand. Indeed, the characteristic gap $\delta\lambda$
between (wanted) states $\ket{\lambda^{\pm}} \simeq (\ket{S} \pm \ket{R})/ \sqrt{2}$
is very 
sensitive to changes in $N$ and to the ratio between weak and strong couplings, say $b$ and $J = J_{\mathrm{max}}$, respectively.
Such behavior emerges very clearly when we set the sequence
$(J, b, J, b, \ldots,b, J)$ for the \textit{channel}. 
(Let us take  $M=N/2$ odd, without loss of generality, only to assure $\Lambda_{N}$ positive.)
Thereby, $\Lambda_{N}$ (and so the effective gap $\delta \lambda = 2J_{\mathrm{eff}}$) will decrease abruptly 
since $\Lambda_{N}\sim b^{M-1}$. 
The above coupling configuration would then improve the fidelity but with 
the cost of having a transfer about  $\Lambda_{N} J$ times slower 
when compared with the uniform channel, $b \rightarrow J$, considering the same $a$.
%

If one decides to \textit{increase} $\Lambda_{N}$ instead, 
say, by 
adjusting the couplings to $(b, J, b, J, \ldots, J, b)$ such that $\Lambda_{N} \sim b^{-M}$,
the transfer time would be about $\Lambda_{N} J$ times faster, but with
unavoidable fidelity loss.
To better see this, let us set $a/J = \xi$ 
for the uniform channel
and $a/J = \xi/\sqrt{\Lambda_{N} J}$ 
for the modified (staggered) channel, with
$\xi \ll 1$. Both channels should now perform 
QST in the same time $\tau \simeq \pi/(2\xi^2 J)$ and we are left 
with the task of comparing the resulting fidelities. 
%

%
\begin{figure}[t!] 
\includegraphics[width=0.4\textwidth]{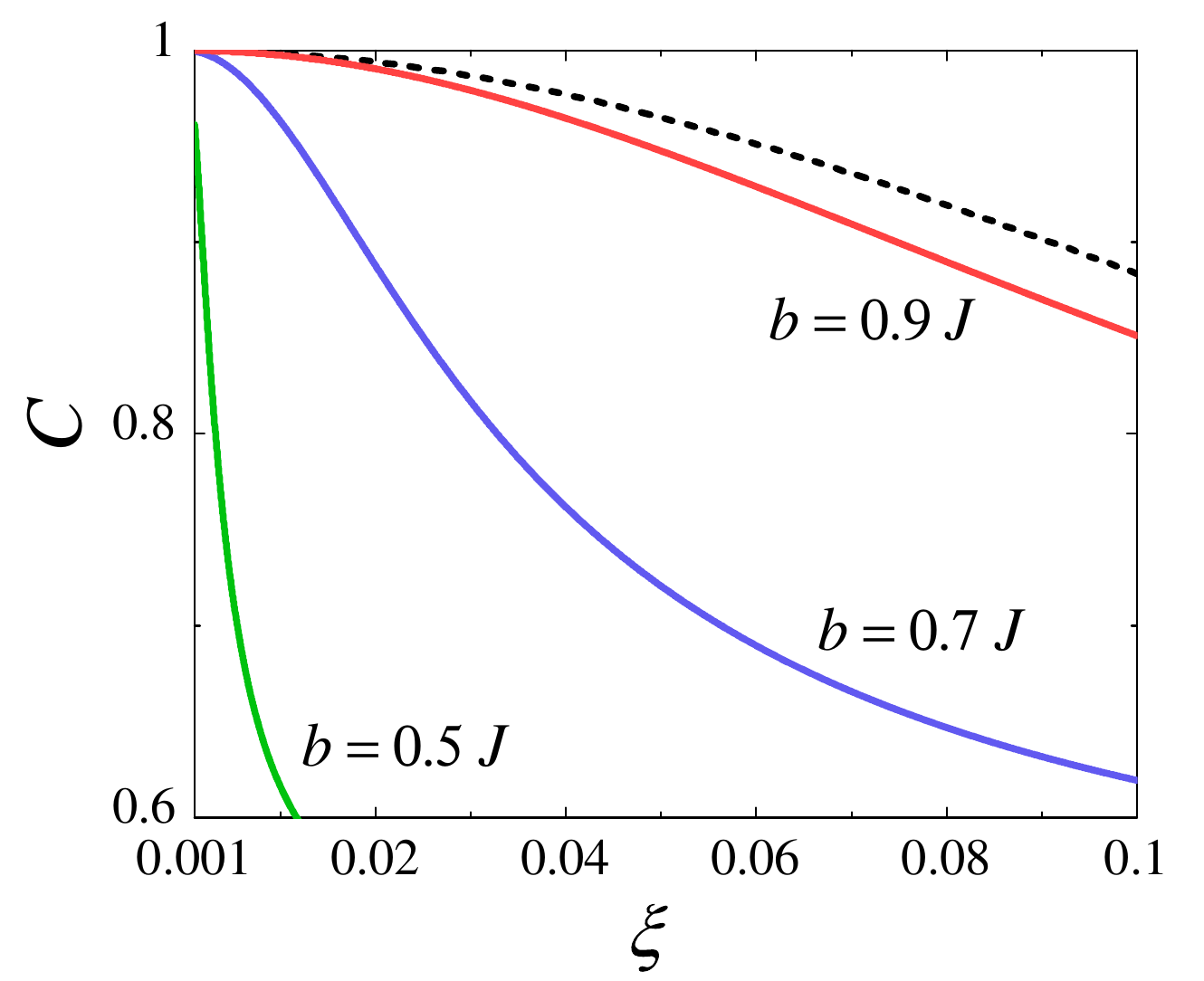}
\caption{\label{fig2} (Color online) End-to-end correlation amplitude $C$ versus $\xi$ (dimensionless parameters) for a channel with $N=30$ featuring the coupling 
pattern $(b, J, b, J, \ldots, J, b)$ for $b/J = 0.9$, $0.7$, $0.5$ (solid lines), 
and $1$ (homogeneous channel; dashed line).
For this setting, $\Lambda_{N} \sim b^{-15} $ [cf. Eq. (\ref{omega})].
Numerical data was obtained from exact numerical diagonalization of Hamiltonian (\ref{Hchannel}) and
by evaluating $C$ for one of the eigenstates closest the band center.
}
\end{figure}

In order to properly operate in the effective Rabi regime, the staggered
channel must satisfy $\xi \sqrt{\Lambda_{N} J} \ll 1$. Therefore, departing from
the uniform channel (for which $\Lambda_{N}= 1/J$), any attempt 
to increase $\Lambda_{N}$ (regardless of the coupling pattern) should 
bring the channel progressively out of the perturbation regime, for fixed $\xi$.
As an example, in Fig. \ref{fig2} we plot the end-to-end correlation amplitude $C$ versus $\xi$ for various 
$\Lambda_{N}$ values, after exact diagonalization of the full Hamiltonian, Eq. (\ref{Hchannel}) for $N=30$ (plus spins $S$ and $R$).
Recall that $C$ primarily indicates how successful the transfer will be, with $C \simeq 1$ implying
$F \simeq 1$  [cf. Eqs. (\ref{eq2}) and (\ref{eq3})]. 
Figure \ref{fig2} clearly shows that the curve for the homogeneous channel bounds
all the other configurations, entailing that if the fidelity is to be evaluated in time window $ \sim O(\xi^{-2})$ -- note that, 
due to the finiteness of $\xi$,
higher-order interactions between sender/receiver and the channel might deviate from $\tau$ a little bit \cite{wojcik05, wojcik07} --
the former always outperforms and, for decreasing $b$ (increasing $\Lambda_{N}$), the decay
is more critical since
the chain now demands smaller $\xi$ values
in order to \textit{secure} the Rabi-like regime.   

From the above analysis we note that for 1D chains with weak end bonds operating in the Rabi-like regime,
given a target value for fidelity $\sim O(1)$, there is
no way to configure the set of couplings of the channel $\lbrace J_{i} \rbrace$
to achieve a faster QST
other than keep them all equal.
Despite we have arrived at this conclusion 
using spatially
symmetric (staggered) channels as an example,
similar arguments hold for non-symmetric channels as well.
%

%
\begin{figure}[t!] 
\includegraphics[width=0.35\textwidth]{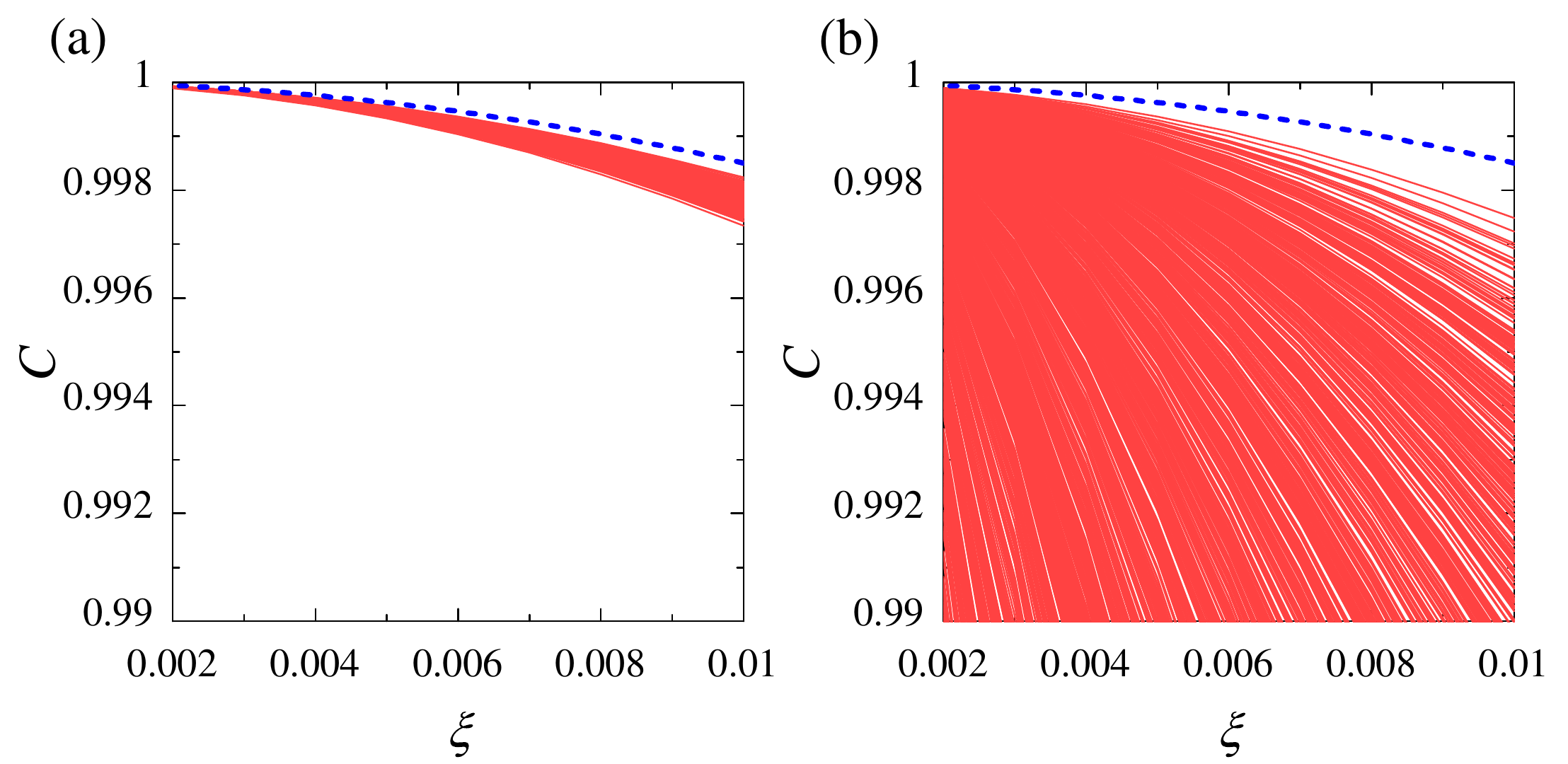}
\caption{\label{fig3} (Color online) End-to-end correlation amplitude $C$ versus $\xi$ (dimensionless parameters) for $N=30$ with 
odd-even paired up random couplings $J_{i} = J_{i+1}$ with the middle one, 
$J_{M} = J_{15}$, \textit{fixed} to $J = J_{\mathrm{max}}$ for $10^{3}$ independent samples (solid red lines)
in comparison with the homogeneous channel configuration (dashed blue line).
The paired couplings were
generated from $J_{i} = J(1-\chi)$, with $\chi$ being a random number uniformly distributed within
$[0,W]$. In (a), $W = 0.5$ and in (b,) $W = 0.99$. 
Note that $\Lambda_{N} = 1/J$ for all realizations [cf. Eq. (\ref{omega})].
Again, numerical data was obtained from exact numerical diagonalization of Hamiltonian (\ref{Hchannel}) and
by evaluating $C$ for one of the eigenstates closest the band center.
}
\end{figure}

Noting in Eq. (\ref{omega}) that $J_{\mu}$ and $J_{\nu}$
with even $\mu$ and odd $\nu$ can cancel each other out,
a question that naturally arises now is whether any other combination of $\lbrace J_{i} \rbrace$ 
resulting in $\Lambda_{N} =1/J =1/J_{\mathrm{max}}$
would outperform the homogeneous channel, which features 
the same property. 
To address this, let us recall our 4-site virtual system
($a$,$\Lambda_{N}^{-1}$,$a$).
Even when we do not have
enough information about the full spectral decomposition of the channel (here reduced to a single
dimer coupled by $\Lambda_{N}^{-1}$), 
we do know -- from Eq. (\ref{omega}) and condition $a\Lambda_{N}\ll 1$ --
its tendency 
to close or open the energy gap in the middle of the band given $\lbrace J_{i} \rbrace$.
If the coupling pattern is such that  $\Lambda_{N} =1/J_{\mathrm{max}}$ and generated 
out of a sequence fulfilling $J_{i} \leq J_{\mathrm{max}}$,
it will shift normal modes towards the band center 
thereby disturbing $\ket{\lambda^{\pm}} \simeq (\ket{S} \pm \ket{R})/ \sqrt{2}$.

As an example, in Fig. \ref{fig3} we show the decay of the end-to-end correlation amplitude
versus $\xi$ for numerous randomly-generated paired up coupling sequences
satisfying $J_{i} = J_{i+1}$, i.e.
$J_{1} = J_{2}$, $J_{3} = J_{4}$, and so forth with exception of $J_{M}$ (again for $M = N/2$ odd without loss of generality)
which has been singled out to $J_{\mathrm{max}}$. It clearly shows 
that none of the samples were able to outperform the regular, uniform channel
in terms of speed-fidelity cost, they all sharing the same $\Lambda_{N} = 1/J_{\mathrm{max}}$.


\section{\label{sec6}Conclusions} 

We have investigated single-qubit QST protocols between two 
parties weakly coupled to the ends of a $XX$ spin-$1/2$ channel with arbitrary (non-symmetric) nearest-neighbor couplings.
We obtained an analytical formula that accounts for the effective coupling between spins $S$ and $R$
given any set of couplings $\lbrace J_{i} \rbrace$. Straightforward analysis
showed that uniform channels, $J_{i} = J$, offer 
the best speed-fidelity cost 
within the Rabi-like (off-resonant) regime.


We would also like to highlight that all the analysis performed in this work was done in the perturbative regime up to second order. 
Naturally, the overall behavior changes considerably as we depart from that scenario. 
For example, a lot of theoretical progress has been achieved for
chains operating in the ballistic regime \cite{banchi10, apollaro12}. 


Finally, our results illuminate the inner machinery Rabi-type QST protocols, 
their physical limitations,
and shall also 
be of interest for studies on long-range entanglement generation schemes  \cite{venuti07,giampaolo09}.
Further extensions of this work could involve non-trivial channel topologies, such as complex networks \cite{almeida13, kostak07,alvarez10}. 
 

%

\section{Acknowledgments}

We thank A. M. C. Souza for valuable discussions.
This work was supported by CNPq (Grant No. 152722/2016-5). 

\appendix*
\section{\label{appendix}Derivation of $\Lambda_{N}$}

In this appendix, we carry out a proof 
of
$\Lambda_{N} = \sum_{k}(v_{k,1}v_{k,N}^{*}/\epsilon_{k})$ as expressed in Eq. (\ref{omega}).
Our starting point is the  
eigenvalue equation
$\hat{H}_{\mathrm{ch}}\ket{\epsilon_{k}} = \epsilon_{k}\ket{\epsilon_{k}}$
for the channel's Hamiltonian $\hat{H}_{\mathrm{ch}}$ [Eq. (\ref{Hchannel})] 
with coupling sequence $(J_{1},J_{2},\ldots,J_{N-1})$ (see Fig. \ref{fig1}).
In matrix form, it reads
\begin{widetext}
\begin{equation}
    \begin{pmatrix}
0 & J_{1} & 0 & \cdots &  &  & 0\\ 
J_{1} & 0 & J_{2} &  &  &  & \\ 
0 & J_{2} & 0 & J_{3} &  &  & \\ 
\vdots &  & J_{3} & 0 &  &  & \vdots\\ 
 &  &  &  & \ddots & J_{N-2} & 0\\ 
 &  &  &  & J_{N-2} & 0 & J_{N-1}\\ 
0 &  &  & \cdots & 0 & J_{N-1} & 0
\end{pmatrix}
\begin{pmatrix}
v_{k,1}\\ 
v_{k,2}\\ 
v_{k,3}\\ 
\vdots\\ 
\\ 
\\ 
v_{k,N}
\end{pmatrix}
=
\epsilon_{k}
\begin{pmatrix}
v_{k,1}\\ 
v_{k,2}\\ 
v_{k,3}\\ 
\vdots\\ 
\\ 
\\ 
v_{k,N}
\end{pmatrix},
\end{equation}
\end{widetext}
where $\lbrace v_{k,i} \rbrace$ and $\lbrace \epsilon_{k} \rbrace$ are all real valued and $N$ is even.
The above relation yields the following set of equations:
\begin{equation} \label{sistema}
    \left\{\begin{matrix}
J_{1}v_{k,2} = \epsilon_{k}v_{k,1}\\ 
J_{1}v_{k,1}+J_{2}v_{k,3} = \epsilon_{k}v_{k,2}\\ 
J_{2}v_{k,2}+J_{3}v_{k,4} = \epsilon_{k}v_{k,3}\\ 
\vdots\\ 
J_{i-1}v_{k,i-1}+J_{i}v_{k,i+1} = \epsilon_{k}v_{k,i}\\ 
\vdots\\ 
J_{N-1}v_{k,N-1} = \epsilon_{k}v_{k,N}
\end{matrix}\right. ,
\end{equation}
from which we get
\begin{equation} \label{v_iplus1}
    \frac{v_{k,i+1}}{\epsilon_{k}} = \frac{v_{k,i}}{J_{i}}-\frac{J_{i-1}v_{k,i-1}}{J_{i}\epsilon_{k}}
\end{equation}

Now, as our sole purpose here is to evaluate the sum $\Lambda = \sum_{k}(v_{k,1}v_{k,N}/\epsilon_{k})$, we may
write it as [cf. Eq. (\ref{v_iplus1})],
\begin{equation} \label{lambda_ana}
  \Lambda_{i+1} = \sum_{k}\frac{v_{k,1}v_{k,i+1}}{\epsilon_{k}} =\frac{1}{J_{1}}\delta_{1,i}-\frac{J_{i-1}}{J_{i}}\sum_{k}\frac{v_{k,1}v_{k,i-1}}{\epsilon_{k}},
\end{equation}
with $i = 1,3,5,\ldots, N-1$ and
we have used $\sum_{k} v_{k,1}v_{k,i} = \delta_{1,i}$ (Kronecker's delta). 
Note that whenever $i=1$ we set the second term on the right side of Eq. (\ref{lambda_ana})
to vanish.

Therefore, we are able to build $\Lambda_{N}$ from $\Lambda_{2} = 1/J_{1}$ --
which is also readily seen from the first relation in Eq. (\ref{sistema}) -- and further iterating it. 
For $N=4$ and $N=6$, say, we get, respectively, $\Lambda_{4} = -J_{2}/(J_{1}J_{3})$ and $\Lambda_{6} = J_{2}J_{4}/(J_{1}J_{3}J_{5})$.
For an arbitrary, \textit{even} $N$ (that is $i=N-1$), we finally get
\begin{align}
  \Lambda_{N} &= \sum_{k}\frac{v_{k,1}v_{k,N}}{\epsilon_{k}} \nonumber \\  &= -\frac{J_{N-2}}{J_{N-1}}\sum_{k}\frac{v_{k,1}v_{k,N-2}}{\epsilon_{k}} \nonumber \\
  &= -\frac{J_{N-2}}{J_{N-1}} \left(-\frac{J_{N-4}}{J_{N-3}} \sum_{k}\frac{v_{k,1}v_{k,N-4}}{\epsilon_{k}} \right) \nonumber \\
  &= -\frac{J_{N-2}}{J_{N-1}} \left(-\frac{J_{N-4}}{J_{N-3}} \right) \cdots  \left( -\frac{J_{N-j}}{J_{N-j-1}} \sum_{k}\frac{v_{k,1}v_{k,N-j}}{\epsilon_{k}} \right) \nonumber \\
  &= \frac{J_{N-2}J_{N-4}\cdots J_{2}}{J_{N-1}J_{N-3}\cdots J_{1}} (-1)^{\frac{N}{2}+1},
\end{align}
which proves Eq. (\ref{omega}).


%

\end{document}